\documentclass[a4paper,11pt]{amsart}%
\usepackage{amsmath, amsfonts, amsthm}
\usepackage{graphicx}

   \DeclareGraphicsExtensions{.eps,.bmp,.png,.jpg}

\begin{document}
\title[Discussion]{Frequent or Systematic Changes? discussion on  ``Detecting
possibly frequent change-points: Wild Binary Segmentation 2 and steepest-drop
model selection.''}

\author{Myung Hwan Seo}
\footnote{This work was supported by the Ministry of Education of the Republic of Korea and the National Research Foundation of Korea (NRF-2018S1A5A2A01033487)}
\address{Department of Economics, Seoul National University, Gwan-Ak Ro 1,
Seoul, Korea. \linebreak myunghseo@snu.ac.kr}

\begin{abstract}
	We discuss  Fryzlewicz's (2020)  that proposes WBS2.SDLL approach to detect possibly frequent changes in mean of a series. Our focus is on the potential issues related to the model misspecification. We present some numerical examples such  as the self-exciting threshold autoregression and the unit root process, that can be confused as a frequent change-points model. 
\end{abstract}

\maketitle

First, I congratulate the author for developing an impressive method called "WBS2.SDLL" to detect frequent change-points in the intercept-only Gaussian regression model within reasonable computing times. I believe this method has also potential to be extended to more general models that can be of greater interest in economics and finance.

Under correct specification, the proposal is undoubtedly valuable addition to the change-point literature. I focus my discussion on the potential misspecification issue in the regression with change-points. Previously, econometric literature has investigated the related issues of confusing the threshold process, unit root process, Markov switching process, and the structural breaks. See for instance, Perrron (1989), Carassco (2002), Seo (2008), among many others. Koo and Seo (2015) also examined the distribution of the change-point estimate when the one-time change-point model is misspecificed from a continuous change model. 

This issue may be more pronounced with the frequent change-points. In the end, one may want a way to distinguish the unpredictable frequent change setup from more systematic predictable changes like cycles or a stochastic trend as in the unit root process.

To illustrate, I ran a small Monte Carlo simulation. Two samples of
size $n=500$ are generated from the following two data generating
processes (dgp):
\begin{align*}
\text{Random Walk: }y_{1t} & =y_{1t-1}+\varepsilon_{t},\\
\text{SETAR(1): }y_{2t} & =\left(a_{1}+b_{1}y_{2,t-1}\right)1\left(y_{2,t-1}>\tau\right)+\varepsilon_{t},
\end{align*}
where $\varepsilon_{t}\sim IIDN\left(0,1\right)$ with $a_{1}=0.7$,
$a_{2}=.7$, $\tau=1$. And the proposed WBS2.SDLL method were applied
to fit the data. A realization of sample path and the fit from the
proposed method is plotted below for each dgp. 

\begin{figure}
	\centering
\caption{A Sample Path from Random Walk and Fit by WBS2.SDLL}
\includegraphics[scale=0.4,angle=270]{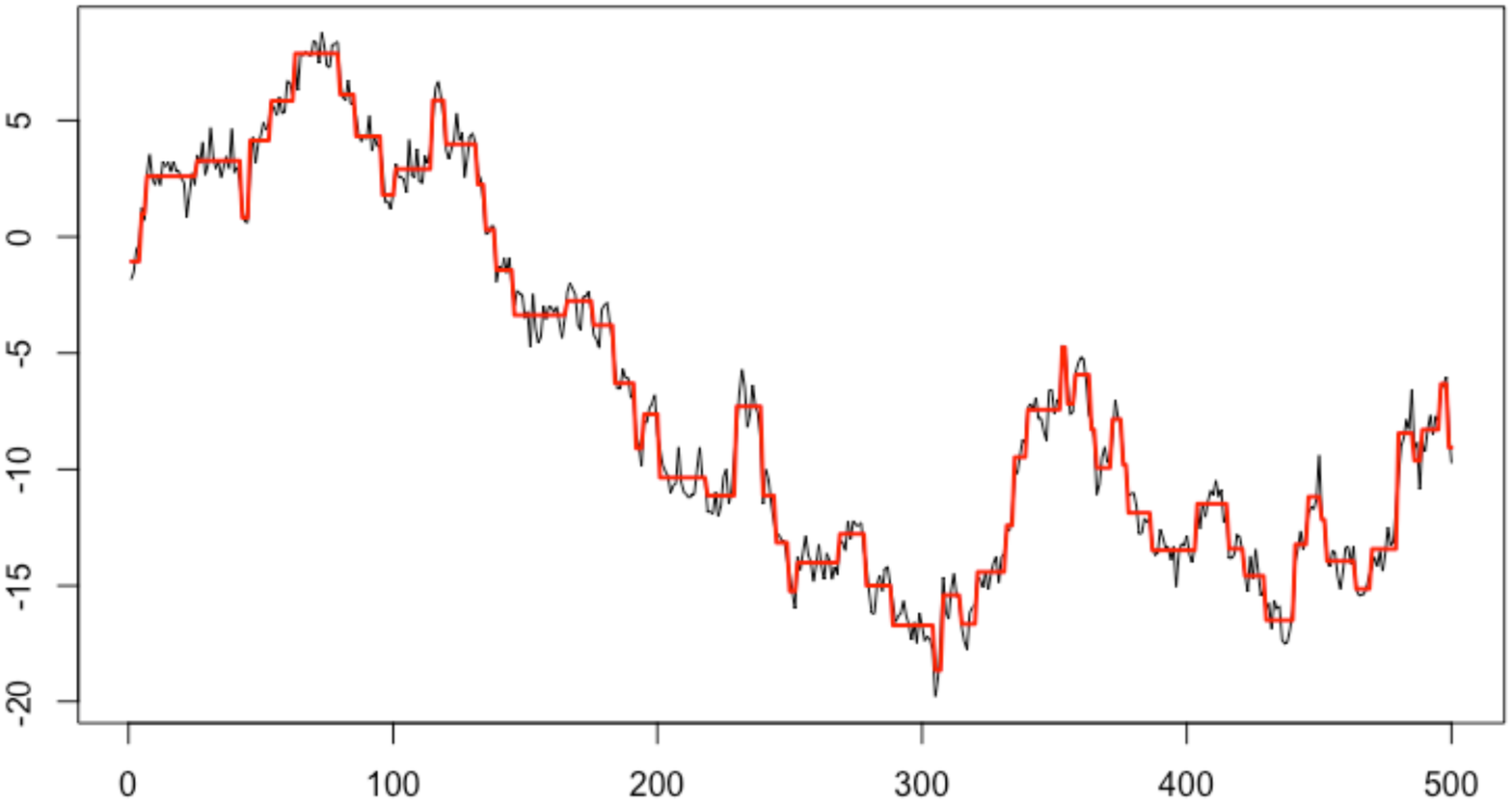}
\end{figure}

\begin{figure}
	\centering
\caption{A Sample Path from SETAR(1) and Fit by WBS2.SDLL}
\includegraphics[scale=0.4,angle=270]{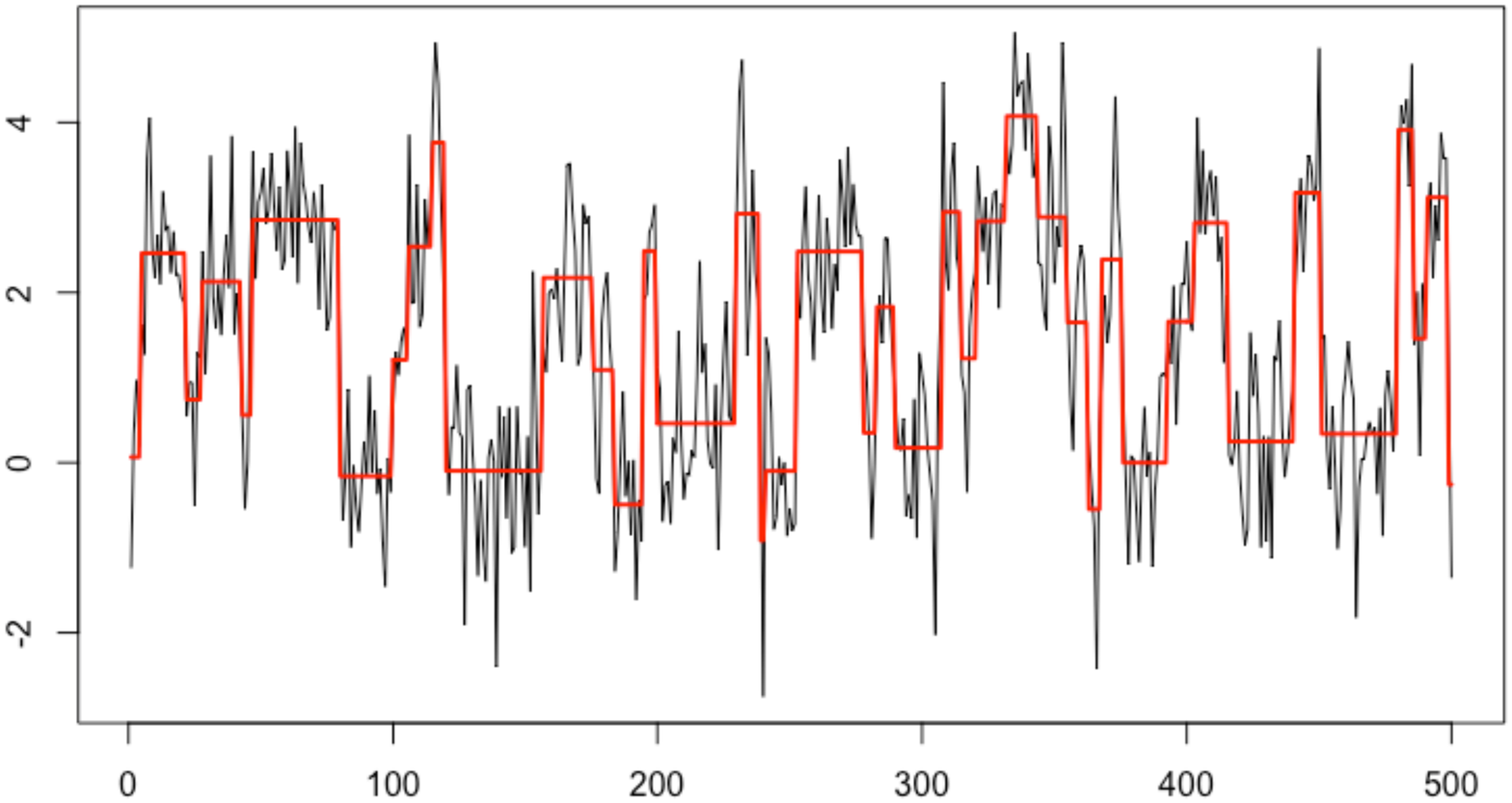}
\end{figure}

The random walk process in Figure 1 exhibits a stochastic trend, while the SETAR process in Figure 2 shows a cyclical movement. The seminal work by Tong and Lim (1980) demonstrated that the SETAR model is capable of capturing various cyclical movement of a time series with a parsimonious parametrization. The above illustrates that the frequent change-points approach via WBS2.SDLL can mistake the systematic movements with a set of many unpredictable change-points. Even Figure 1 in the author's paper can also be viewed as a recurring predictable changes or cycle. 

In further investigation, I repeated the experiment 200 times. The WBS2.SDLL yielded an average of 67.1 (s.e. 45.7) change-points for the random walk $y_{1t}$ and 20.4 (s.e.18.9) change-points for the SETAR process $y_{2t}$. When one has detected frequent changes, tyring to understand the nature of the frequent changes will be very important and deserves more attention.


\begin{thebibliography}{1}
\bibitem{key-1}Carrasco, M. (2002). Misspecified structural change,
threshold, and Markov-switching models. Journal of econometrics, 109(2),
239-273.

\bibitem{} Fryzlewicz, P. (2020). Detecting possibly frequent change-points: Wild Binary Segmentation 2 and steepest-drop model selection. Journal of the Korean Statistical Society,
https://doi.org/10.1007/s42952-020-00060-x

\bibitem{key-6}Koo, B., \& Seo, M. H. (2015). Structural-break models
under mis-specification: Implications for forecasting. Journal of
econometrics, 188(1), 166-181.

\bibitem{key-4}Perron, P. (1989). The great crash, the oil price
shock, and the unit root hypothesis. Econometrica: journal of the
Econometric Society, 1361-1401.

\bibitem{key-3}Seo, M. H. (2008). Unit root test in a threshold autoregression:
asymptotic theory and residual-based block bootstrap. Econometric
Theory, 24(6), 1699-1716.

\bibitem{key-2}Tong, H., \& Lim, K. S. (1980). Threshold autoregression,
limit cycles and cyclical data-with discussion. Journal of the Royal
Statistical Society. Series B: Statistical Methodology, 42(3), 245-292.
\end{thebibliography}
\end{document}